\documentclass[final,3p,times]{elsarticle}
\usepackage{amssymb}
\usepackage{amsmath}
\usepackage{multicol}
\usepackage{graphicx}
\usepackage{graphics}
\usepackage{epsfig}
\usepackage{dcolumn}
\usepackage{bm}
\usepackage{threeparttable}
\newcommand{\thickhline}{\noalign{\hrule height 0.8pt}}

\journal{Physica B}
\begin{document}
\begin{frontmatter}
\title{Thickness dependent Curie temperature and power-law behavior of layering transitions in ferromagnetic classical and quantum 
thin films described by Ising, XY and Heisenberg models}
\author[]{Yusuf Y\"{u}ksel\corref{cor1}}
\cortext[cor1]{Corresponding author. Tel.: +90 2323019547; fax: +90 2324534188.}
\ead{yusuf.yuksel@deu.edu.tr}
\author{\"{U}mit Ak{\i}nc{\i}}
\address{Dokuz Eyl\"{u}l University, Department of Physics, T{\i}naztepe Campus, TR-35160 Izmir, Turkey}
\begin{abstract}
Ferromagnetic-paramagnetic phase transitions in classical and quantum thin films have been studied up to 50 mono-layers 
using effective field theory with two-site cluster approximation. Variation of the Curie temperature as a function of film thickness has been examined.
The relative shift of the Curie temperature from the corresponding bulk value has been investigated in terms of the shift exponent $\lambda$. We have found that
shift exponent $\lambda$ clearly depends on the strength of the
ferromagnetic exchange coupling of the surface. Moreover, we have
not observed any significant difference between classical and quantum exponents for a particular model.
\end{abstract}
\end{frontmatter}
\section{Introduction}\label{introduction}
By the development of modern and sensitive growth techniques such as ultra-high vacuum and molecular beam epitaxy, characterization and realization of thin magnetic
films became experimentally accessible even in the mono-layer limit which consequently led to considerable amount of interest in thin film magnetism in the last four decades, both from
theoretical and experimental points of view \cite{pleimling,poupoulos,vaz,kaneyoshi1}. In contrast to the bulk magnetism, surface effects which mainly originate as a
result of reduced coordination number of the surface of the material are prominent in real physical systems. The existence of surfaces simply causes a broken symmetry
in the system. Therefore, magnetic properties of finite magnetic materials such as magnetic thin films may differ from those of bulk materials. As a consequence
of these facts, the surface may exhibit an ordered phase even if the bulk itself is disordered which has already been experimentally observed \cite{ran,polak,tang}.

From the theoretical point of view, the models based on Ising-type spin Hamiltonians successfully explain the critical phenomena of highly anisotropic
magnetic uniaxial thin films \cite{strandburg} by utilizing well known theoretical tools including Monte Carlo (MC) simulations \cite{tucker},
mean field theory (MFT) \cite{aguilera}, and effective field theory (EFT) \cite{kaneyoshi2}. However, a vast majority of magnetic materials in nature do not exhibit
such a strong uniaxial anisotropy. Hence, the theoretical investigation of magnetism in these systems require more realistic models. In this context,
Heisenberg model on thin film geometry has been studied by a wide variety of methods such as renormalization group (RG) method \cite{mariz}, high
temperature series expansions (HTSE) \cite{masrour,masrour2,masrour3,pan,ritchie}, density functional theory (DFT) \cite{hortamani}, EFT
\cite{neto0,neto1,neto2,akinci1,akinci2,benyoussef,neto3,neto6,wu}, MFT \cite{neto7,neto8,jensen}, and Green functions formalism (GFF)
\cite{diep1,diep2,diep3,selzer,corciovei}. EFT method which is superior to conventional MFT is widely used in the literature. For instance, Neto and co-workers utilized this method
to investigate classical and quantum Heisenberg thin films \cite{neto0,neto1,neto2,neto3,neto6}. One of their most remarkable findings is that the N\'{e}el temperature
$T_{N}$ of an anti-ferromagnetic (AF) quantum Heisenberg film is greater than Curie temperature $T_{c}$ of a ferromagnetic (FM) quantum Heisenberg film 
whereas in the classical case the two models (i.e. AF and FM) exhibit the same critical temperature values. In principle, in order to apply the EFT formulation for quantum and classical Heisenberg models, one should improve the standard EFT method \cite{kaneyoshi0} by
defining the spin-spin interactions on a larger cluster which was formerly introduced by Bob\'{a}k et al. \cite{bobak} and known as EFT-2
formulation in the literature \cite{balcerzak,albuquerque0, fittipaldi0,fittipaldi1,sousa0,sousa1,albuquerque1,sousa2,araujo,idogaki0,idogaki1}.

On the other hand, theoretical and experimental investigations are also focused on the finite size shift of the critical temperature $T_{c}(L)$
of the film from the corresponding bulk critical temperature as a function of film thickness which is characterized by a shift exponent.
More clearly, for a thin film with thickness $L$, it follows a power law behavior \cite{baberschke0} for thick films
\begin{equation}
\varepsilon=1-\frac{T_{c}(L)}{T_{c}(\infty)}\propto L^{-\lambda},
\end{equation}
where $\lambda$ is the shift exponent, and it is directly related to the bulk correlation length exponent as $\nu_{b}=1/\lambda$ \cite{barber}. High resolution
MC simulations yield that for 3D Ising and Heisenberg models, the exponent values are $\lambda=1.588$ \cite{ferrenberg0} and
$\lambda=1.419$ \cite{chen}, respectively. The universality behavior of Ising thin films has been examined widely in the literature \cite{yuksel}.

Experimentally, Bramfeld and Willis \cite{bramfeld} reported results for nickel films and they observed a
thickness dependent dimensional crossover. Fuchs et al. \cite{fuchs0,fuchs1} investigated the variation of magnetization as a function of film thickness
for lanthanum films and extracted an exponent value $\lambda=0.9$ which resembles a value obtained by MFT for Ising model. For ultra-thin nickel films, a shift
exponent $\lambda=1.7$ was found \cite{tizliouine}, and also a crossover from a three- to a two-dimensional magnetic behavior was observed by  Li et al. \cite{li}. For
thin iron films \cite{henkel}, phenomenological finite-size scaling analysis yields an effective shift exponent $\lambda=3.15$ which is twice as large as the value
expected from the conventional finite-size scaling prediction \cite{barber} whereas for amorphous iron and aluminum films \cite{korelis} $\lambda=1.2$ were found.  	
Furthermore, thickness dependent crossover of the exponent $\lambda$ was also observed for gadolinium films in the formalism of finite size scaling \cite{waldfried}.

Although the topic has attracted a considerable amount of interest, the dimensionality in the thin films is not well established and the investigation of
universality properties, especially for thicker films, needs particular attention \cite{neto2}. Hence, in the present paper, our aim is to investigate
the thickness dependent Curie temperature and its relative shift from the bulk critical point using EFT-2 formulation for ferromagnetic thin films described
by classical and quantum spin models. The paper can be outlined as follows: In Sec. \ref{formulation} we give a brief formulation of EFT-2 method for thin film systems.
Sec. \ref{results} is devoted to numerical results and discussions, and finally Sec. \ref{conclusions} contains our conclusions.

\section{Formulation}\label{formulation}

In this work, we consider a magnetic thin film with $L$ successive layers (see Fig. \ref{fig1}). Classical and quantum versions of the Hamiltonian
describing our model can be written as
\begin{subequations}
\begin{eqnarray}\label{eq1a}
\mathcal{H}^{\mathrm{Classical}}&=&-\sum_{<ij>}J_{ij}\mathbf{S}_{i}.\mathbf{S}_{j},\\
\label{eq1b}\mathcal{H}^{\mathrm{Quantum}}&=&-\sum_{<ij>}J_{ij}(\delta_{x}\sigma_{i}^{x}\sigma_{j}^{x}
+\delta_{y}\sigma_{i}^{y}\sigma_{j}^{y}+\delta_{z}\sigma_{i}^{z}\sigma_{j}^{z}).
\end{eqnarray}
\end{subequations}
In classical model described by Eq. (\ref{eq1a}), $\mathbf{S}_{i}$ is a $d-$ dimensional classical spin vector where $d=1,2,3$ corresponds to Ising, classical XY, and
classical Heisenberg systems, respectively. On the other hand, for the quantum model described by Eq. (\ref{eq1b}), $\delta_{\alpha}$ is the anisotropy parameter,
and $\sigma_{i}^{\alpha}$ represents the $\alpha-$ component of the Pauli spin operator with $\alpha=x,y,z$. For $\delta_{x,y}=0$ and $\delta_z=1$, we recover
Ising Hamiltonian whereas for $\delta_{x}=0$ and $\delta_{y,z}=1$ we have quantum XY model, and for $\delta_{x,y,z}=1$, the model corresponds
to isotropic quantum Heisenberg system. $J_{ij}$ in each Hamiltonian stands for the exchange couplings between nearest-neighbor spins. Namely, if the two spins are located at the surface region of the film then $J_{ij}=J_{s}$
otherwise, we have $J_{ij}=J_{b}$ (c.f. Fig. \ref{fig1}). Following the same methodology given in previous works \cite{albuquerque0} and utilizing the EFT-2 formulation \cite{bobak} with Kaneyoshi-Honmura differential
operator technique and decoupling approximation \cite{kaneyoshi0,honmura}, we obtain the longitudinal components of the layer magnetizations as
\begin{eqnarray}\label{eq2}
\nonumber
M_{1}&=&[\cosh(K_{s}\nabla_{x})+M_{1}\sinh(K_{s}\nabla_{x})]^{z_{0}-1}[\cosh(K_{s}\nabla_{y})+M_{1}\sinh(K_{s}\nabla_{y})]^{z_{0}-1}\\
\nonumber
&&[\cosh(K_{b}\nabla_{x})+M_{2}\sinh(K_{b}\nabla_{x})][\cosh(K_{b}\nabla_{y})+M_{2}\sinh(K_{b}\nabla_{y})]g_{s}(x,y)|_{x,y=0},\\
\nonumber&&\\
\nonumber
M_{\nu}&=&[\cosh(K_{b}\nabla_{x})+M_{\nu}\sinh(K_{b}\nabla_{x})]^{z_{0}-1}[\cosh(K_{b}\nabla_{y})+M_{\nu}\sinh(K_{b}\nabla_{y})]^{z_{0}-1}\\
\nonumber
&&[\cosh(K_{b}\nabla_{x})+M_{\nu+1}\sinh(K_{b}\nabla_{x})][\cosh(K_{b}\nabla_{y})+M_{\nu+1}\sinh(K_{b}\nabla_{y})]\\
\nonumber
&&[\cosh(K_{b}\nabla_{x})+M_{\nu-1}\sinh(K_{b}\nabla_{x})][\cosh(K_{b}\nabla_{y})+M_{\nu-1}\sinh(K_{b}\nabla_{y})]g_{b}(x,y)|_{x,y=0},\\
\nonumber&&\\
\nonumber
M_{L}&=&[\cosh(K_{s}\nabla_{x})+M_{L}\sinh(K_{s}\nabla_{x})]^{z_{0}-1}[\cosh(K_{s}\nabla_{y})+M_{L}\sinh(K_{s}\nabla_{y})]^{z_{0}-1}\\
&&[\cosh(K_{b}\nabla_{x})+M_{L-1}\sinh(K_{b}\nabla_{x})][\cosh(K_{b}\nabla_{y})+M_{L-1}\sinh(K_{b}\nabla_{y})]g_{s}(x,y)|_{x,y=0},
\end{eqnarray}
where $2\leq\nu\leq L-1$ is the layer index, and the terms $K_{s}$ and $K_{b}$ are respectively defined as $K_{s}=\beta J_{s}$ and $K_{b}=\beta J_{b}$
with $\beta=1/k_{B}T$. For simplicity, we set $k_{B}=1$. The term $z_{0}$ in Eq. (\ref{eq2}) is the intra-layer coordination number. Since we are interested in
thin films with simple cubic structure, we have $z_{0}=4$. For the classical thin film, the functions $g_{s}(x,y)$ and $g_{b}(x,y)$ in Eq. (\ref{eq2}) 
are of the form \cite{neto0,neto2,albuquerque1,stanley}
\begin{subequations}
 \begin{equation}\label{eq3}
g_{\alpha}(x,y)=\frac{\sinh(x+y)}{\cosh(x+y)+\exp(-2K_{\alpha})\phi(K_{\alpha}d)\cosh(x-y)},
\end{equation}
with
\begin{equation}\label{eq4}
\phi(K_{\alpha}d)=\frac{I_{d/2-1}(K_{\alpha}d)-I_{d/2}(K_{\alpha}d)}{I_{d/2-1}(K_{\alpha}d)+I_{d/2}(K_{\alpha}d)}\exp(2K_{\alpha}), \quad \alpha=b \ \mathrm{or} \ s
\end{equation}
\end{subequations}
where $I_{n}(x)$ is the modified Bessel function of the first kind.

For the quantum case, we have \cite{idogaki1}
\begin{subequations}
\begin{equation}\label{eq3_ek}
g_{\alpha}(x,y)=\frac{(x+y)\delta_{z}}{X_{0}}\frac{\sinh X_{0}}{\cosh X_{0}+\exp(-2K_{\alpha}\delta_{z})\cosh Y_{0}},
\end{equation}
with
\begin{equation}\label{eq4_ek}
X_{0}=\sqrt{(x+y)^{2}\delta_{z}^{2}+(\delta_x-\delta_y)^{2}K_{\alpha}^{2}}, \quad
Y_{0}=\sqrt{(x-y)^{2}\delta_{z}^{2}+(\delta_x+\delta_y)^{2}K_{\alpha}^{2}}.
\end{equation}
\end{subequations}

In order to proceed further, we apply binomial expansion
\begin{equation}\label{eq5}
(x+y)^{n}=\sum_{i=0}^{n}
\left(\begin{tabular}{c}
  $n$  \\
  $i$  \\
\end{tabular}\right)
x^{n-i}y^{i},
\end{equation}
in Eq. (\ref{eq2}). After some mathematical manipulations, we obtain
\begin{eqnarray}\label{eq6}
\nonumber
M_{1}&=&\sum_{i=0}^{z_{0}-1}\sum_{j=0}^{z_{0}-1}\sum_{k=0}^{1}\sum_{l=0}^{1}\lambda_{s}(i,j,k,l)M_{1}^{i+j}M_{2}^{k+l},\\
\nonumber
M_{\nu}&=&\sum_{i=0}^{z_{0}-1}\sum_{j=0}^{z_{0}-1}\sum_{k=0}^{1}\sum_{l=0}^{1}\sum_{m=0}^{1}\sum_{n=0}^{1}
\lambda_{b}(i,j,k,l,m,n)M_{\nu}^{i+j}M_{\nu+1}^{k+l}M_{\nu-1}^{m+n},\\
\nonumber
M_{L}&=&\sum_{i=0}^{z_{0}-1}\sum_{j=0}^{z_{0}-1}\sum_{k=0}^{1}\sum_{l=0}^{1}\lambda_{s}(i,j,k,l)M_{L}^{i+j}M_{L-1}^{k+l},\\
\end{eqnarray}
where
\begin{eqnarray}\label{eq7}
\nonumber
\lambda_{s}(i,j,k,l)&=&
\left(\begin{tabular}{c}
$z_{0}-1$  \\
$i$  \\
\end{tabular}\right)
\left(\begin{tabular}{c}
$z_{0}-1$  \\
$j$  \\
\end{tabular}\right)
\Theta_{s}(i,j,k,l)g_{s}(x,y)|_{x,y=0},\\
\lambda_{b}(i,j,k,l,m,n)&=&
\left(\begin{tabular}{c}
$z_{0}-1$  \\
$i$  \\
\end{tabular}\right)
\left(\begin{tabular}{c}
$z_{0}-1$  \\
$j$  \\
\end{tabular}\right)
\Theta_{b}(i,j,k,l,m,n)g_{b}(x,y)|_{x,y=0},
\end{eqnarray}
with
{\small
\begin{eqnarray}\label{eq8}
\nonumber
\Theta_{s}(i,j,k,l)&=&2^{-2z_{0}}\sum_{p_{1}=0}^{(z_{0}-1-i)}\sum_{p_{2}=0}^{(z_{0}-1-j)}
\sum_{p_{3}=0}^{1-k}\sum_{p_{4}=0}^{1-l}\sum_{p_{5}=0}^{i}\sum_{p_{6}=0}^{j}
\sum_{p_{7}=0}^{k}\sum_{p_{8}=0}^{l}(-1)^{(p_{5}+p_{6}+p_{7}+p_{8})}\\
\nonumber
&&\times
\left(\begin{tabular}{c}
$z_{0}-1-i$  \\
$p_{1}$  \\
\end{tabular}\right)
\left(\begin{tabular}{c}
$z_{0}-1-j$  \\
$p_{2}$  \\
\end{tabular}\right)
\left(\begin{tabular}{c}
$1-k$  \\
$p_{3}$  \\
\end{tabular}\right)
\left(\begin{tabular}{c}
$1-l$  \\
$p_{4}$  \\
\end{tabular}\right)
\left(\begin{tabular}{c}
$i$  \\
$p_{5}$  \\
\end{tabular}\right)
\left(\begin{tabular}{c}
$j$  \\
$p_{6}$  \\
\end{tabular}\right)
\left(\begin{tabular}{c}
$k$  \\
$p_{7}$  \\
\end{tabular}\right)
\left(\begin{tabular}{c}
$l$  \\
$p_{8}$  \\
\end{tabular}\right)\\
\nonumber
&&\times\exp[\{K_{s}(z_{0}-1-2p_{1}-2p_{5})+K_{b}(1-2p_{3}-2p_{7})\}\nabla_{x}]\\
\nonumber
&&\times\exp[\{K_{s}(z_{0}-1-2p_{2}-2p_{6})+K_{b}(1-2p_{4}-2p_{8})\}\nabla_{y}],\\
\nonumber
\Theta_{b}(i,j,k,l,m,n)&=&2^{-(2z_{0}+2)}
\sum_{p_{1}=0}^{(z_{0}-i-k-m+1)}
\sum_{p_{2}=0}^{(z_{0}-j-l-n+1)}
\sum_{p_{3}=0}^{(i+k+m)}
\sum_{p_{4}=0}^{(j+l+n)}
(-1)^{(p_{3}+p_{4})}\\
\nonumber
&&\times
\left(\begin{tabular}{c}
$z_{0}-i-k-m+1$  \\
$p_{1}$  \\
\end{tabular}\right)
\left(\begin{tabular}{c}
$z_{0}-j-l-n+1$  \\
$p_{2}$  \\
\end{tabular}\right)
\left(\begin{tabular}{c}
$i+k+m$  \\
$p_{3}$  \\
\end{tabular}\right)
\left(\begin{tabular}{c}
$j+l+n$  \\
$p_{4}$  \\
\end{tabular}\right)\\
&&\times\exp[K_{b}(z_{0}-2p_{1}-2p_{3}+1)\nabla_{x}+K_{b}(z_{0}-2p_{2}-2p_{4}+1)\nabla_{y}].
\end{eqnarray}}

By solving Eqs. (\ref{eq7}) and (\ref{eq8}) numerically using the identity
$\exp[(\alpha_{x}\nabla_{x})+(\alpha_{y}\nabla_{y})]g_{b,s}(x,y)=g_{b,s}(x+\alpha_{x},y+\alpha_{y})|_{x,y=0}$
for any given set of system parameters, and using them in Eq. (\ref{eq6}), we obtain a
system of non linear equations which contains the polynomial forms of layer magnetizations. The longitudinal magnetization $M_{i}$ of each layer can be obtained
from numerical solution of Eq. (\ref{eq6}).

On the other hand, in order to obtain the transition temperature of a film with a particular thickness $L$ and given system parameters, one should linearize
Eq. (\ref{eq6}), hence we obtain
\begin{eqnarray}\label{eq9}
\nonumber
M_{1}&=&[\lambda_{s}(1,0,0,0)+\lambda_{s}(0,1,0,0)]M_{1}+[\lambda_{s}(0,0,1,0)+\lambda_{s}(0,0,0,1)]M_{2},\\
\nonumber
&& \\
\nonumber
M_{\nu}&=&[\lambda_{b}(1,0,0,0,0,0)+\lambda_{b}(0,1,0,0,0,0)]M_{\nu}+[\lambda_{b}(0,0,1,0,0,0)+\lambda_{b}(0,0,0,1,0,0)]M_{\nu+1}\\
\nonumber
&&+[\lambda_{b}(0,0,0,0,1,0)+\lambda_{b}(0,0,0,0,0,1)]M_{\nu-1},\\
\nonumber
&& \\
M_{L}&=&[\lambda_{s}(1,0,0,0)+\lambda_{s}(0,1,0,0)]M_{L}+[\lambda_{s}(0,0,1,0)+\lambda_{s}(0,0,0,1)]M_{L-1},
\end{eqnarray}
where $2\leq\nu\leq L-1$. Using Eq. (\ref{eq9}), the transition temperature can be numerically evaluated by solving $\mathrm{det}(A)=0$ where $A$ is the coefficient
matrix of the linearized equations in Eq. (\ref{eq9}). From many possible solutions of the condition $\mathrm{det}(A)=0$, we have to choose the highest possible
one corresponding to the transition temperature of the system.

\section{Results and discussion}\label{results}
In order to provide a testing ground for our calculations, the typical phase diagrams of classical and quantum XY films in a $(T_{c}/J_{b}-J_{s}/J_{b})$ plane
are plotted in Fig. \ref{fig2} which exhibit a phenomenon peculiar to magnetic thin films. Namely, in a system with a surface, the phase diagrams corresponding
to different film thickness $L$
intersect each other at a special point which can be denoted by $(T_{c}(\infty)/J_{b},J_{s}^{*}/J_{b})$ where $T_{c}(\infty)/J_{b}$ is the reduced transition temperature of
the corresponding bulk system (i.e. simple cubic lattice in this case) and $J_{s}^{*}/J_{b}$ is the critical value of the surface to bulk ratio of exchange
interactions
of the film. For classical and quantum XY thin films, the locations of the special points are $(5.034,1.321)$ and $(4.980,1.332)$,
respectively which was previously reported in
Ref. \cite{neto1}. Based on Fig. \ref{fig2}, it is a well known fact that the thicker films have higher critical values for $J_{s}/J_{b}<J_{s}^{*}/J_{b}$
corresponding to the ordinary case
while for extraordinary case which is characterized by $J_{s}/J_{b}>J_{s}^{*}/J_{b}$, the thicker films exhibit lower transition
temperatures than the thinner ones.

For the investigation of the thickness dependent Curie temperature and its relative shift from the bulk value, the exponent $\lambda$ can be extracted from
numerical data by plotting $\varepsilon$ versus $L$ curves for sufficiently thick films in a log-log scale then fitting the resultant curve using the
standard linear regression method. In order to precisely cover the critical region, the obtained data have been fitted for those providing the condition
$0.001\leq\varepsilon\leq0.01$ which generally requires to consider the transition temperatures of the films with $L>20$ in fitting procedure. The illustrative
results are demonstrated in Fig. \ref{fig3}  with $J_{s}/J_{b}=1.0$ and up to $L=50$ mono-layers for classical and quantum Heisenberg thin films, respectively.
Our calculations yield the exponent values $\lambda=1.909$ and $\lambda=1.907$ for the classical and quantum Heisenberg thin films, respectively. The results are also
summarized in Table-\ref{table1} for several thin film models and in the presence of modified surfaces.

According to Table-\ref{table1}, our results show the trend $\lambda(I)>\lambda(XY)>\lambda(H)$ which agrees well with those obtained by HTSE calculations \cite{masrour3} and
MFT predictions \cite{neto8}.
It is also clear that for the Ising thin films, the obtained values using EFT-2 are clearly greater than those obtained
by EFT-1 formulation \cite{yuksel}. Apart from these,
one can conclude from Table-\ref{table1} that the presence of modified surface exchange interactions clearly affect the value of the exponent $\lambda$.
Namely, for the weak surface couplings such as $J_{s}/J_{b}=0.5$, quantum exponents are barely larger than the classical exponents whereas 
in the presence of a moderate surface coupling such as $J_{s}/J_{b}=1.0$, an opposite scenario takes place. Moreover, it seems like a dimensional crossover 
may take place as $J_{s}/J_{b}$ varies. In other words, the exponent value $\lambda$ is very close to the bulk value for small $J_{s}/J_{b}$. This behavior can be
attributed to three-dimensional character of the film, since $\nu_{b}=1/\lambda=0.5$ \cite{barber} whereas for greater values of $J_{s}/J_{b}$, the system tends to
reveal a two dimensional character even for thicker films.
It is also worth to  note that our exponent values for the Heisenberg thin films agree well with those obtained using GFF \cite{corciovei} in which $\lambda(H)=2.0$
was predicted. However, HTSE \cite{masrour3,pan,ritchie} and MFT \cite{neto8} yields rather small exponents in comparison to EFT-2 results which may be due to
the fact that the thickness of the films considered in these works lacks to provide the condition $0.001\leq\varepsilon\leq0.01$.
\begin{table}[h]
\begin{center}
\begin{threeparttable}
\caption{Extracted exponent values for Ising, XY and Heisenberg thin films in the presence of modified surfaces.}
\label{table1}
\renewcommand{\arraystretch}{1.5}
\begin{tabular}{cccccc}
\thickhline
& Ising & Classical & Quantum & Classical & Quantum  \\
&       &  XY &   XY &  Heisenberg & Heisenberg  \\
\hline
$J_{s}/J_{b}=0.5$ \ \ \  & 2.035 & 2.011 & 2.017 & 2.004  & 2.007 \\
$J_{s}/J_{b}=1.0$ \ \ \  & 1.955 & 1.920 & 1.915 & 1.909  & 1.907 \\
\thickhline \\
\end{tabular}
\end{threeparttable}
\end{center}
\end{table}

\section{Conclusions}\label{conclusions}
In this work, by utilizing EFT-2 method, we have examined the phase diagrams and universality properties of ferromagnetic classical and quantum thin films described by Ising, XY and Heisenberg models. Analysis for the thickness dependent Curie temperature and its relative shift from the bulk value reveals that the value of the shift exponent $\lambda$ clearly depends on the strength of the 
ferromagnetic exchange coupling of the surface. In case of Ising thin films, the obtained values of $\lambda$ using EFT-2 are found to be clearly larger than those obtained
by EFT-1 formulation. Both in the classical and quantum pictures, the trend $\lambda(I)>\lambda(XY)>\lambda(H)$ agrees well with those obtained by other methods in the literature. 
However, although the critical properties such as bulk transition temperature $T_{c}(\infty)$ differ apparently between classical and quantum formulations, we have
 not observed any significant difference between classical and quantum exponents for a particular model (i.e. XY or Heisenberg).


\newpage
\textbf{Figure Captions}

Fig.1 Schematic representation of a ferromagnetic thin film.

Fig.2 Variation of the transition temperature as a function of modified
surface exchange interactions for a magnetic thin film with various thickness $L$ described by XY model. Filled circle represents
the location of the special point. (a) classical XY, (b) quantum XY models.

Fig.3 Variation of the shift exponent $\lambda$ with surface to bulk ratio of exchange couplings $J_{s}/J_{b}=1.0$ for (a) classical and (b) quantum Heisenberg
thin films.


\setcounter{section}{0}
\setcounter{figure}{0}
\newpage

\begin{figure}[h!]
\center
\includegraphics[width=11.0cm]{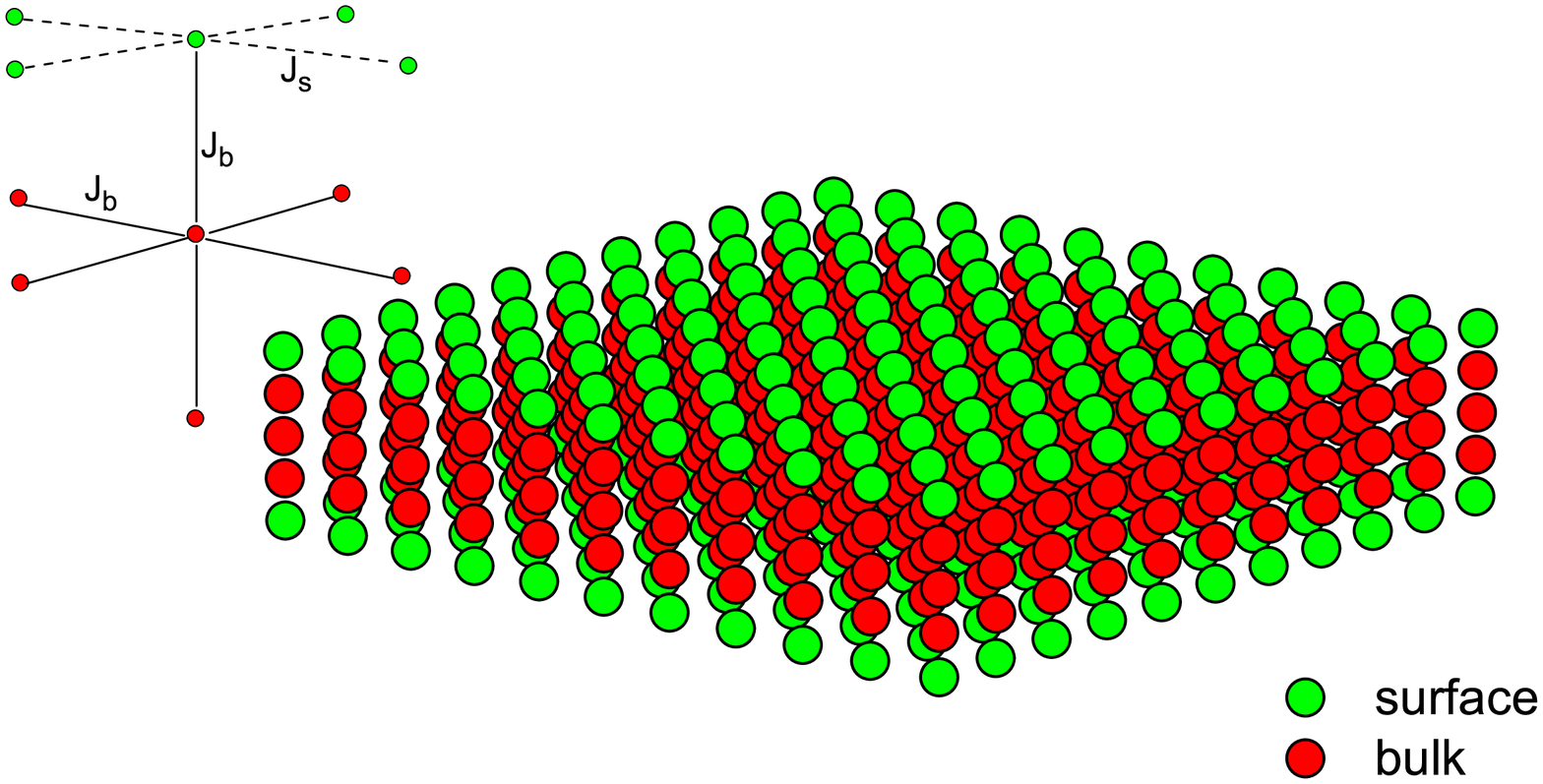}
\caption{}
\label{fig1}
\end{figure}

\newpage
\begin{figure}[h!]
\includegraphics[width=8.0cm]{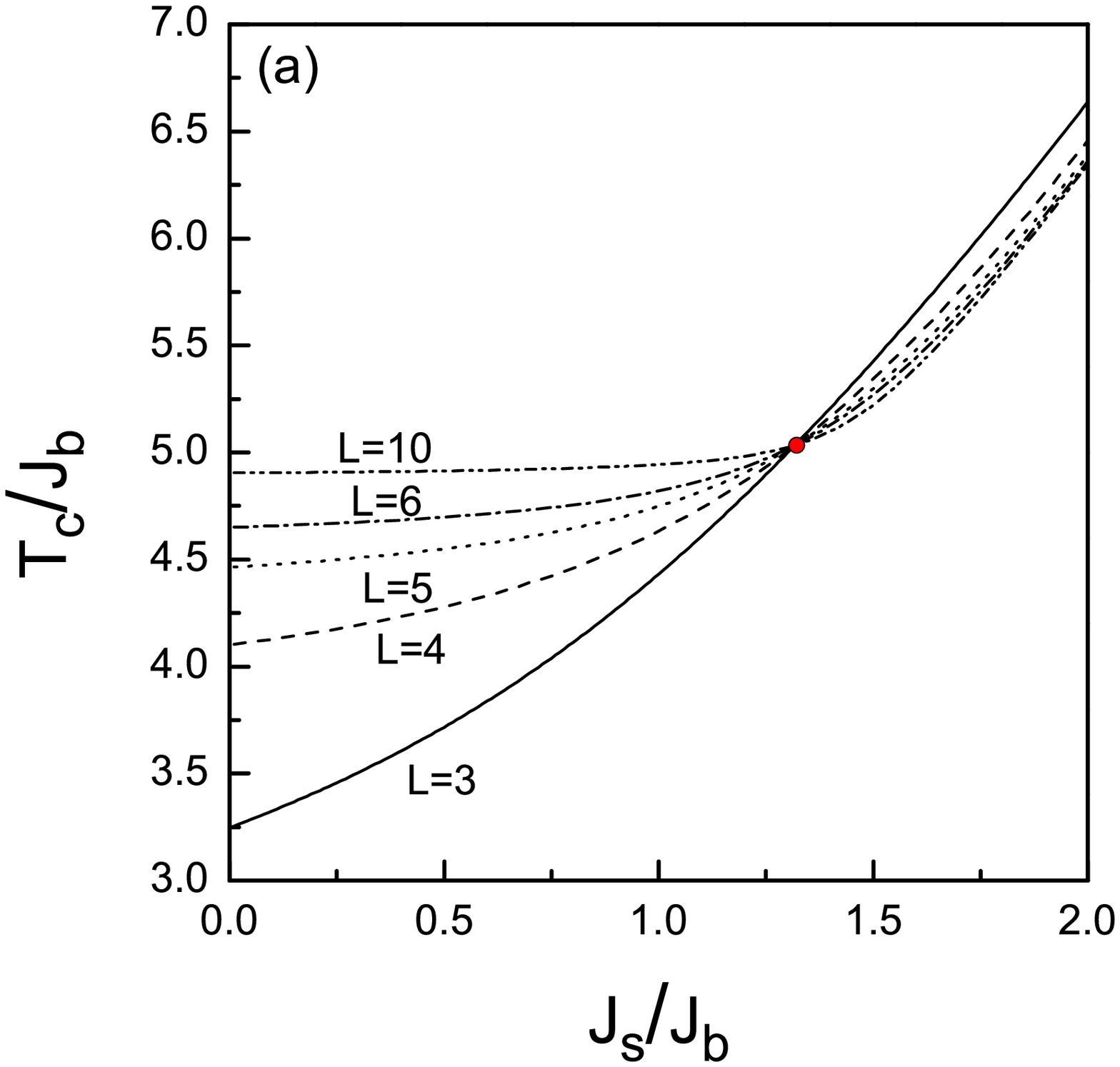}
\includegraphics[width=8.0cm]{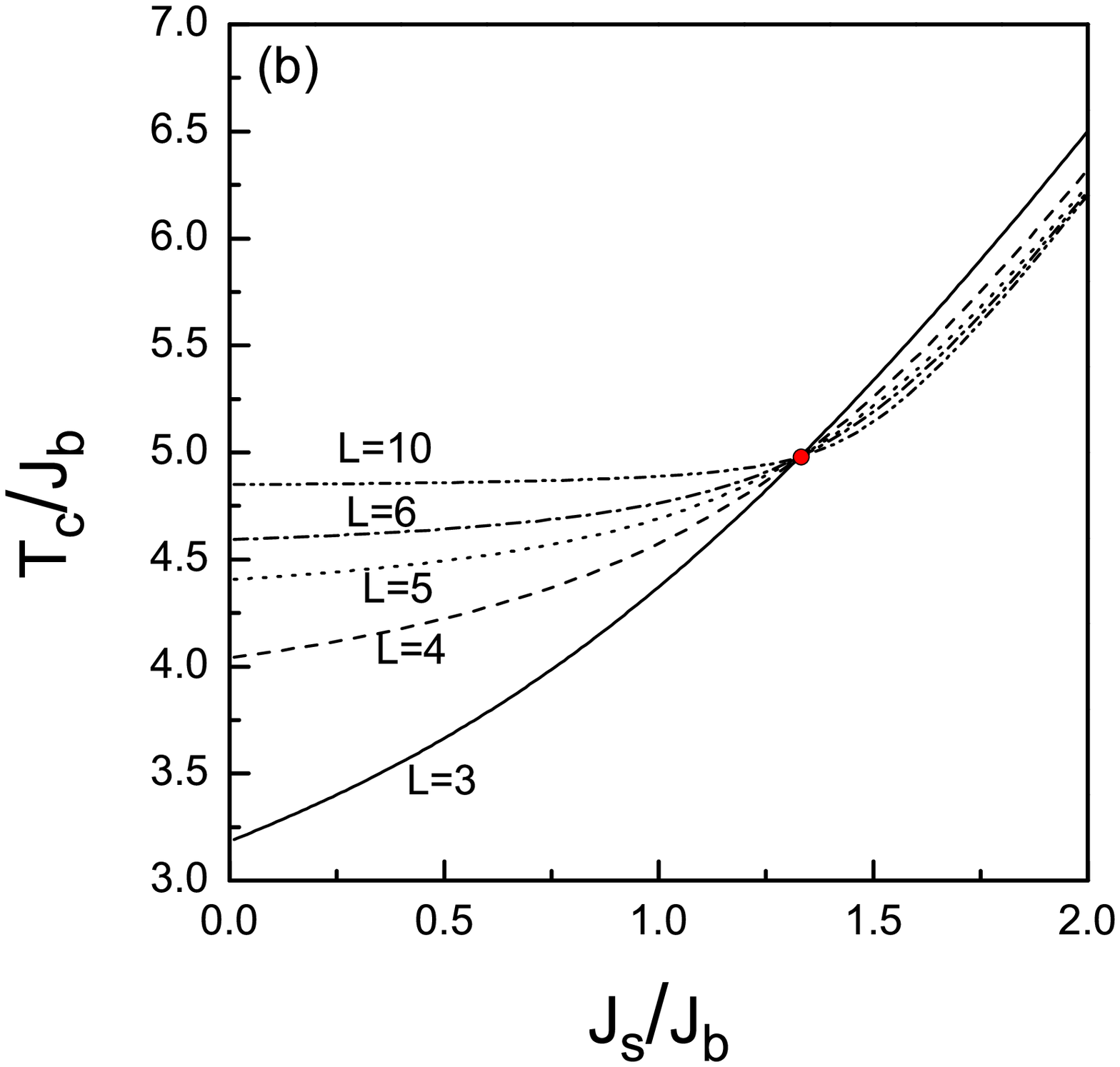}\\
\caption{}
\label{fig2}
\end{figure}

\newpage
\begin{figure}[h!]
\includegraphics[width=8.0cm]{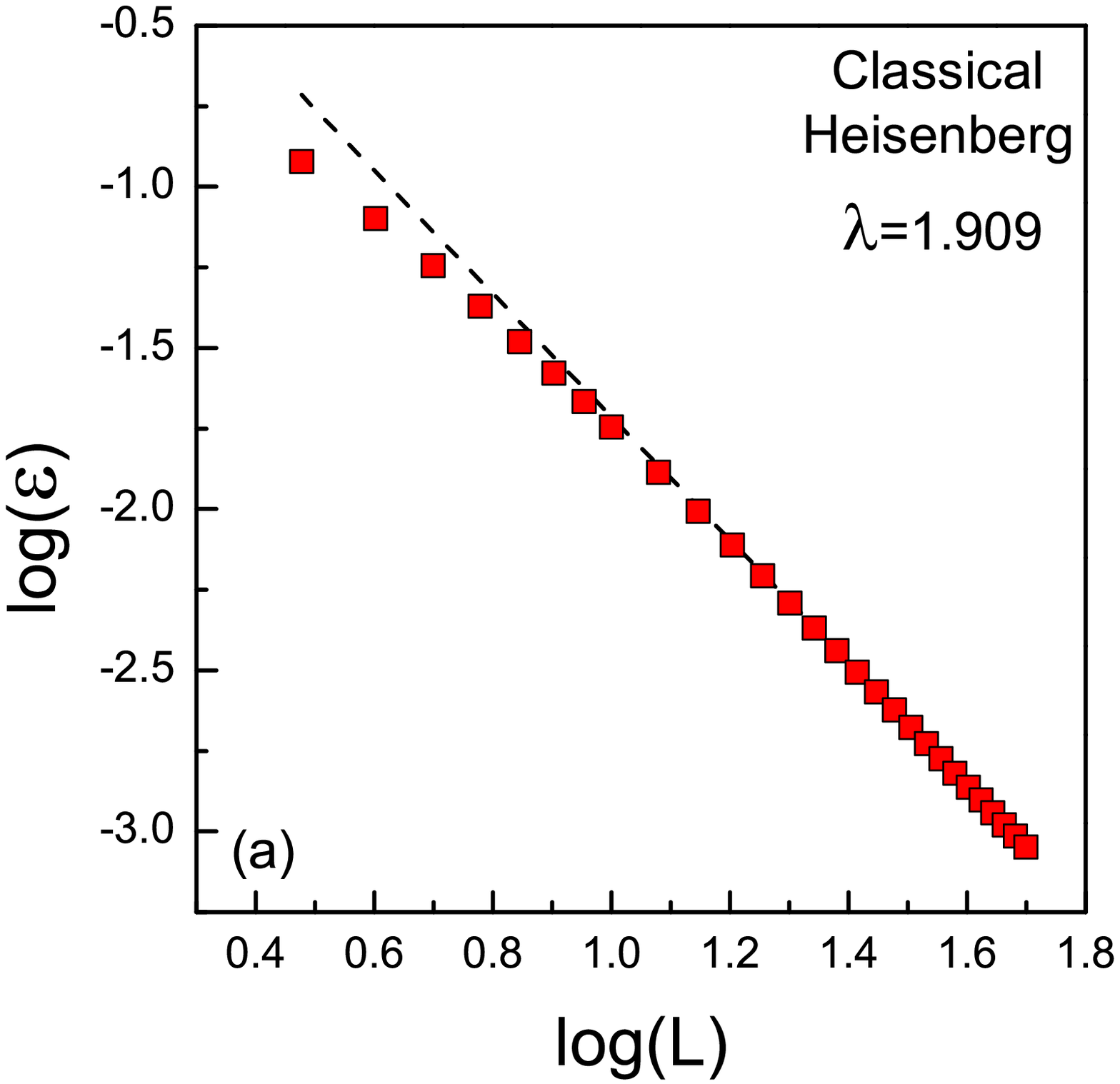}
\includegraphics[width=8.0cm]{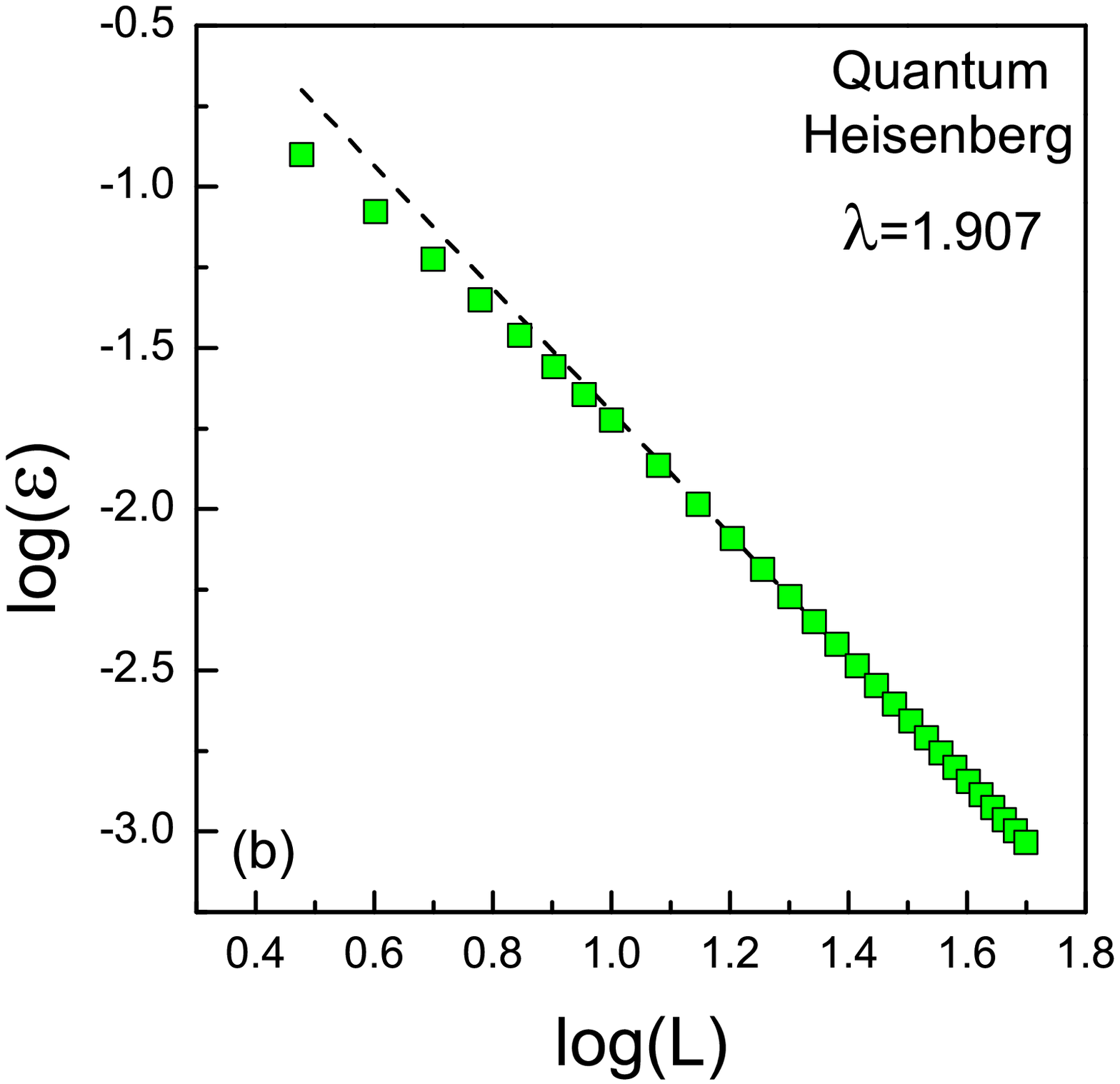}\\
\caption{}
\label{fig3}
\end{figure}


\begin{thebibliography}{99}
\bibitem{pleimling} M. Pleimling, Critical phenomena at perfect and non-perfect
surfaces, J. Phys. A: Math. Gen. 37 (2004) R79.
\bibitem{poupoulos} P. Poupoulos, K. Baberschke, Magnetism in thin films, J. Phys.: Condens. Matter 11 (1999) 9495.
\bibitem{vaz} C.A.F. Vaz, J.A.C. Bland, G. Lauhoff, Magnetism in ultrathin film structures, Rep. Prog. Phys. 71 (2008) 056501.
\bibitem{kaneyoshi1} T. Kaneyoshi, Surface magnetism; magnetization and anisotropy at a surface, J. Phys.: Condens. Matter 3 (1991) 4497.
\bibitem{ran} C. Ran, C. Jin, M. Roberts, Ferromagnetic order at Tb surfaces above the bulk Curie temperature, J. Appl. Phys. 63 (1988) 3667.
\bibitem{polak} M. Polak, L. Rubinovich, J. Deng, Observation of highly enhanced Curie temperature at Ni-Al alloy surfaces, Phys. Rev. Lett. 74 (1995) 4059.
\bibitem{tang} H. Tang, Magnetic reconstruction of the Gd(0001) surface, Phys. Rev. Lett. 71 (1985) 444.

\bibitem{strandburg} K. J. Strandburg, D. W. Hall, C. Liu, S. D. Bader, Monte Carlo simulations of the Curie temperature of ultrathin ferromagnetic films, Phys. Rev. B 46 (1992) 10818.

\bibitem{tucker} J.W. Tucker, A Monte Carlo study of thin spin-1 Ising films with surface exchange enhancement, J. Magn. Magn. Mater. 210 (2000) 383.
\bibitem{aguilera} F. Aguilera-Granja, J.L. Mor\'{a}n L\'{o}pez, Ising model of phase transitions in ultrathin films, Solid State Commun. 74 (1990) 155.
\bibitem{kaneyoshi2} T. Kaneyoshi, Phase diagram of Ising thin films with decorated surfaces; decoupling approximation, Physica A 293 (2001) 200.
\bibitem{mariz} A. M. Mariz, U. M. S. Costa, C. Tsallis, Influence of the interaction anisotropy on the appearence of
surface magnetism, Eur. Phys. Let. 3 (1987) 27.
\bibitem{masrour} R. Masrour, M. Hamedoun, A. Benyoussef, Study of critical behaviour in diluted ferromagnetic: thin films and semi-infinites films,  
Int. J. Mod. Phys B 24 (2010) 3561.
\bibitem{masrour2} R. Masrour, M. Hamedoun, A. Hourmatallah, K. Bouslykhane, N. Benzakour, Magnetic properties of a ferromagnet spin-S, Ising, XY and Heisenberg 
models semi-infinites systems, Phys. Lett. A 372 (2008) 5203.
\bibitem{masrour3} R. Masrour, M. Hamedoun, A. Benyoussef, Phase transition in Ising, XY and Heisenberg magnetic films, Appl. Surf. Sci. 258 (2012) 1902.
\bibitem{pan} K. K. Pan, Magnetic phase transition in Heisenberg antiferromagnetic films with
easy-axis single-ion anisotropy, Physica A 391 (2012) 1984.
\bibitem{ritchie} D. S. Ritchie, M. E. Fisher, Finite-size and surface effects in Heisenberg films, Phys. Rev. B 7 (1973) 480.
\bibitem{hortamani} M. Hortamani, L.M. Sandratskii, I. Mertig, Does a Heisenberg Hamiltonian describe magnetic interactions
in a MnSi film properly?, J. Magn. Magn. Mater. 322 (2010) 1082.
\bibitem{neto0} J. C. Neto, J. R. de Sousa, Phase diagrams of thin classical n-vector films, J. Magn. Magn. Mater. 268 (2004) 298.
\bibitem{neto1} J. C. Neto, J. R. de Sousa, Surface magnetism: phase transitions in quantum
and classical models, Physica A 319 (2003) 319.
\bibitem{neto2} J. C. Neto, J. R. de Sousa, J. A. Plascak, Critical properties of thin quantum and classical Heisenberg films, Phys. Rev. B 66 (2002) 06447.
\bibitem{akinci1} \"{U}. Ak{\i}nc{\i}, Random field distributed Heisenberg model on a thin film geometry, J. Magn. Magn. Mater. 368 (2014) 36.
\bibitem{akinci2} \"{U}. Ak{\i}nc{\i}, Anisotropic Heisenberg model in thin film geometry, Thin Solid Films, 550 (2014) 602.
\bibitem{benyoussef} A. Benyoussef, A. Boubekri, H. Ez-Zahraouy, M. Saber, The Semi-infinite anisotropic
spin-i Heisenberg ferromagnet, Chin. J. Phys. 37 (1999) 89.
\bibitem{neto3} J. C. Neto, J. R. de Sousa, Magnetic properties of a thin quantum spin-1/2 Heisenberg film, Physica A 364 (2006) 223.
\bibitem{neto6} J. C. Neto, J. R. de Sousa, Phase diagram of a thin Heisenberg antiferromagnetic film, Phys. Status Solidi B 244 (2007) 3361.
\bibitem{wu} Y. Z. Wu, Z. Y. Li, Surface  magnetism  in  the  semi-infinite  Heisenberg  ferromagnet, Solid Stat. Commun. 106 (1998) 789.
\bibitem{neto7} J. C. Neto, J. R. de Sousa, Study of surface effects in quantum Heisenberg
and XY antiferromagnets, Phys. Status Solidi B 212 (1999) 343.
\bibitem{neto8} M. S. Amazonas, J. C. Neto, J. R. de Sousa, Finite size scaling and universality class in ultrathin films, J. Magn. Magn. Mater. 270 (2004) 119.
\bibitem{jensen} P. J. Jensen, H. Dreysse, K. H. Bennemann, Thickness  dependence  of  the  magnetization  and the  Curie
temperature  of  ferromagnetic  thin  films, Surf. Sci. 269-270 (1992) 627.
\bibitem{diep1} H. T. Diep, Quantum effects in Heisenberg antiferromagnetic thin films, Phys. Rev. B 43 (1991) 8509.
\bibitem{diep2} H. T. Diep, J. C. S. Levy, O. Nagai, Effects of surfacc spin waves and surface anisotropy
in magnetic thin films at finite temperatures, Phys. Status Solidi B 93 (1979) 351.
\bibitem{diep3} H. T. Diep, Temperature-dependent surface magnetization
and critical temperature of ferromagnetic thin films, Phys. Status Solidi B 103 (1981) 809.
\bibitem{selzer} S. Selzer, N. Majlis, Effects of surface exchange anisotropy in Heisenberg ferromagnetic insulators, Phys. Rev. B 27 (1983) 544.
\bibitem{corciovei} A. Corciovei, D. Mazilu, D. Mihalache, Quadrupolar-paramagnetic critical temperature
of a spin-one isotropic Heisenberg film with biquadratic interactions, Phys. Status Solidi B 87 (1978) 61.

\bibitem{kaneyoshi0} T. Kaneyoshi, Differential operator technique in the Ising spin systems, Acta Phys. Pol. A 83 (1993) 703.
\bibitem{honmura} R. Honmura, T. Kaneyoshi, Contribution to the new type of effective-field theory of the Ising model, J. Phys. C 12 (1979) 3979.

\bibitem{bobak} A. Bob\'{a}k, M. Ja\v{s}\v{c}ur, A new type of effective field theory for Ising model with spin-1/2, Phys. Status Solidi B 135 (1986) K9.

\bibitem{balcerzak} T. Balcerzak, A. Bob\'{a}k, M. Ja\v{s}\v{c}ur, J. Mielnicki, G. Wiatrowski, Two-atom cluster approximation for Ising model (S=1/2), Phys. Status Solidi B 143 (1987) 261.
\bibitem{albuquerque0} D. F. de Albuquerque, I. P. Fittipaldi, A  unified  effective-field  renormalization-group framework approach for the
quenched  diluted  Ising  models, J. Appl. Phys. 75 (1994) 5832.
\bibitem{fittipaldi0}  I. P. Fittipaldi, D. F. de Albuquerque, Effective-field  renormalization-group  method  for  Ising  systems, J. Magn. Magn. Mater. 104-107 (1992) 236.
\bibitem{fittipaldi1}  I. P. Fittipaldi, Effective  field  renormalization  group  approach for  Ising  lattice  spin  systems, J. Magn. Magn. Mater. 131 (1994) 43.
\bibitem{sousa0} J. R. de Sousa, D. F. de Albuquerque, Critical  properties  of  the  classical  XY  and
classical  Heisenberg  models: a  renormalization  group  study, Physica A 236 (1997) 419.
\bibitem{sousa1} J. R. de Sousa, Equivalence of the O(n) vector ferromagnetic and antiferromagnetic models, Physica A 256 (1998) 383.
\bibitem{albuquerque1} D. F. de Albuquerque, Behavior critical for bond diluted n-vector model
in the effective field theory, Physica A 287 (2000) 185.
\bibitem{sousa2} J. R. de Sousa, J. A. Plascak, Phase transitions in the classical n-vector model on the fcc lattice, Phys. Rev. B 77 (2008) 024419.
\bibitem{araujo} I. G. Ara\'{u}jo, J. C. Neto, J. R. de Sousa, N\'{e}el temperature and sublattice magnetiztion of a
three-dimensional classical and quantum Heisenberg antiferromagnet, Physica A 260 (1998) 150.
\bibitem{idogaki0} T. Idogaki, Y. Miyoshi, J. W. Tucker, An  effective  field  theory  for  dilute  anisotropic  Heisenberg ferromagnets, J. Magn. Magn. Mater. 154 (1996) 221.
\bibitem{idogaki1} T. Idogaki, N. Uryu, A  new  effective  field  theory  for  the  anisotropic Heisenberg  ferromagnet, Physica A 181 (1992) 173.
\bibitem{baberschke0} K. Baberschke, The magnetism of nickel monolayers, Appl. Phys. A: Mater. Sci. Process. 62 (1996) 417.

\bibitem{barber} M. N. Barber, Finite-size scaling, in: J. L. Lebowitz (Ed.), Phase Transitions and Critical Phenomena,
vol. 8, Academic, London, 1983, pp 146.

\bibitem{ferrenberg0} A. M. Ferrenberg, D. P. Landau, Critical hehavior of the three-dimensional Ising model: A high-resolution Monte Carlo study, Phys. Rev. B 44 (1991) 5081.
\bibitem{chen} K. Chen, A. M. Ferrenberg, D. P. Landau, Static critical behavior of three-dimensional classical Heisenberg models:
A high-resolution Monte Carlo study, Phys. Rev. B 48 (1993) 3249.
\bibitem{yuksel} Y. Y\"{u}ksel, \"{U}. Ak{\i}nc{\i}, Universality aspects of layering transitions in ferromagnetic
Blume-Capel thin films, Physica B 433 (2014) 96, and the references therein.

\bibitem{bramfeld} T. S. Bramfeld, R. F. Willis, Temperature-dependent crossover of dimensionality
in ultrathin nickel films, J. Appl. Phys.  103 (2008) 07C718.
\bibitem{fuchs0} D. Fuchs, O. Mor\'{a}n, P. Adelmann, R. Schneider, Thickness-dependent Curie temperature of epitaxial $\mathrm{La_{0.7}A_{0.3}CoO_{3}}$ (A=Ca, Sr, Ba) films, 
J. Magn. Magn. Mater.  272-276 (2004) e1377.
\bibitem{fuchs1} D. Fuchs, T. Schwarz, O. Mor\'{a}n, P. Schweiss, R. Schneider, Finite-size shift of the Curie temperature of ferromagnetic lanthanum cobaltite thin films,   
Phys. Rev. B 71 (2005) 092406.
\bibitem{tizliouine} A. Tizliouine, H. Salhi, H. Lassri, N. Addou, Magnetic studies of spin wave excitations in Ni/Cu multilayers,  
J. Mod. Phys. 2 (2011) 1285

\bibitem{li} Y. Li, K. Baberschke, Dimensional crossover in ultrathin Ni(111) films on W(110), Phys. Rev. Lett. 68 (1992) 1208.
\bibitem{henkel} M. Henkel, S. Andrieu, P. Bauer, M. Piecuch, Finite-size scaling in thin Fe/Ir(100) layers, Phys. Rev. Lett. 80 (1998) 4783.
\bibitem{korelis} P. T. Korelis, P. E. Jonsson, A. Liebig, H. E. Wannberg, P. Nordblad, B. Hjorvarsson, 
Finite-size effects in amorphous $\mathrm{Fe_{90}Zr_{10}/Al_{75}Zr_{25}}$ multilayers, Phys. Rev. B 85 (2012) 214430
\bibitem{waldfried} C. Waldfried, D. Welipitiya, T. McAvoy, P. A. Dowben, Finite size scaling in the thin film limit, J. Appl. Phys. 83 (1998) 7246.
\bibitem{stanley} H. E. Stanley, Exact solution for a linear chain of isotropically interacting classical spins of arbitrary dimensionality, Phys. Rev. 179 (1969) 570.
 

\end{thebibliography}
\end{document}